\documentclass[a4]{epl2}
\usepackage{graphicx}
\usepackage{epsfig}
\usepackage{color}
\usepackage{amsmath}
\usepackage{float}
\title{\textcolor[rgb]{0.00,0.00,0.00}{Coupling of a biquaternionic Dirac field to a bosonic field}}

\author{A. I. Arbab\inst{}\footnote{arbab.ibrahim@gmail.com, aiarbab@uofk.edu}}

\institute{
   \inst{}Department of Physics, College of Science, Qassim University, P.O. Box 6644, 51452 Buraidah,  Kingdom of Saudi Arabia\\
   Department of Physics,
Faculty of Science, University of Khartoum, P.O. Box 321, Khartoum 11115, Sudan}

\abstract{We extend the biquaternionic Dirac equation to include interactions with a background bosonic field. The
obtained biquaternionic Dirac equation yields Maxwell-like equations that hold for both a matter field and
an electromagnetic field. We establish that the electric field is perpendicular to the matter magnetic field
and the magnetic field is perpendicular to the matter inertial field. We show that the inertial and magnetic
masses are conserved separately. The magnetic mass density arises as a result of the coupling between
the vector potential and the matter inertial field. The presence of the vector and scalar potentials and
also the matter inertial and magnetic fields modify the standard form of the derived Maxwell equations.
The resulting interacting electrodynamics equations are generalizations of the equations of Wilczek or
Chern–Simons axion-like fields. The coupled field in the biquaternioic Dirac field reconstructs the Wilczek
axion field. We show that the electromagnetic field vector $\vec{F} =\vec{E} +ic\vec{B}$, where $\vec{E}$ and $\vec{B}$ are the respective
electric and magnetic fields, satisfies the massive Dirac equation and, moreover, $\vec{\nabla}\cdot\vec{F}=0$.}

\begin{document}
\maketitle
\baselineskip=2\baselineskip
\section{\textcolor[rgb]{0.00,0.07,1.00}{Introduction}}

The interaction of photons with an electron is coined in the so called quantum electrodynamics (QED). This is formally done by replacing the ordinary partial derivative in Dirac's equation by the covariant derivative involving the electromagnetic vector and scalar potentials, $\vec{A}$ and $\varphi$ \textcolor[rgb]{0.00,0.07,1.00}{\cite{griff}}. Consequently, Dirac's equation is found to predict the spin of the electron.

In a recent paper we formulated  quantum mechanics in which Dirac's equation is expressed in biquaternionic form \textcolor[rgb]{0.00,0.07,1.00}{\cite{quaternion}}. This representation is found to allow Dirac's equation to be expressible in Maxwell-like equations. In this description, the electron is described by vector and scalar wavefunctions, \emph{viz.}, $\vec{\psi}$ and $\psi_0$ \textcolor[rgb]{0.00,0.07,1.00}{\cite{analogy}}. There are two fields associated with the mass of the particle. They are functions of $\vec{\psi}$ and $\psi_0$ in an analogous way to the electric and magnetic fields that are associated with the particle charge.  We call these two fields the \emph{inertial}  and \emph{magnetoelectrical} fields, $\vec{E}_m$ and $\vec{B}_m$.  In this new formulation, we have found that quantum mechanics can be visualized as a theory for the evolution of matter fields, in analogy with Maxwell theory that describes the evolution of charge fields.

In the present work, we study the interaction of matter fields with an external electromagnetic field (photon) \textcolor[rgb]{0.00,0.07,1.00}{\cite{analogy}}. To this end, we follow a routine that is analogous to that done for quantum electrodynamics. In doing so, we have found that the biquaternionic Dirac's equation can be described by a symmetrized Maxwell-like equations involving magnetic mass as well.

The concept of magnetic mass has to be presumptively interpreted in the framework of this formulation. Both inertial mass and magnetic mass are separately conserved. We treat the mass in quantum mechanics as the charge is treated in electrodynamics. Mass waves are de Broglie waves whereas charge waves are electromagnetic waves. Therefore, they are otherwise analogous to each other. The resulting electrodynamic equations reduce to axions electrodynamics of Frank Wilczek and Chert-Simon \textcolor[rgb]{0.00,0.07,1.00}{\cite{chern1, chern2}}. It was realized by Weinberg  and Wilczek  that axions resulted from CP violation in Quantum-Chromodynamics (QCD) \textcolor[rgb]{0.00,0.07,1.00}{\cite{weinberg,chern1}}.

In free space the resulting electrodynamic equations are found to be invariant under  a duality transformation. This electrodynamics is also invariant under parity and time-reversal transformations. While Wilczek's equations are classical, our present equations are of quantum nature. The electrodynamics we have obtained generalizes Wilczek axions electrodynamics to include magnetic charge and current densities. Moreover, our vacuum electrodynamics is invariant under duality transformation.

This paper proceeds as follows:  In Section 2 we present the biquaternionic interacting Dirac's equation. We adopt here the standard approach as  done for  QED. The resulting equations are that of Maxwellian form with interacting additional fields that are related to the particle inertia (mass)). These additional fields are independent of the energy conservation equation associated with  the electromagnetic fields. The  mass fields, current  are defined in terms of the fundamental constants and the wave functions. From these quantities Maxwell-like equations relating the mass fields are written. The additional fields render the charge (mass) to be conserved.  In Section 2 we discuss the symmetries reflected in these equations. As in the original Maxwell's equations, the resulting Maxwell's equations are found to be invariant under duality transformation in vacuum.

 In Section 3 we present a comparison between Wilczek axion electrodynamics and our electrodynamics. The comparison reveals that axion electrodynamics emerges as a special case. We define a continuity equation that restores the charge conservation by adding a London's-like current to the electric current. Our electrodynamics reduces to Maxwell's equations in a moving frame for special choice of the interacting fields.  In Section 4 we how that our electrodynamics can be expressed in a Dirac-like equation if we employ spin-1 matrices, and writing the electric and magnetic fields in a complex form. We end the paper with  concluding remarks.

\section{\textcolor[rgb]{0.00,0.07,1.00}{The interacting biquaternionic Dirac's equation}}

The ordinary Dirac's equation of a spin-1/2 particle with rest mass $m$ and charge $q$ interacting with a gauge field is expressed as \textcolor[rgb]{0.00,0.07,1.00}{\cite{bjorken}}
\begin{equation}\label{1}
(p^\mu-qA^\mu)\gamma_\mu \psi=mc\,\psi\,,
\end{equation}
where $\gamma^\mu$ are the Dirac's matrices that are expressed in terms of Pauli matrices ($\vec{\sigma})$, $c$ is the speed of light,  $\psi$ are the spinor representing the Dirac's particle, $q$ and $m$ are the particle charge and  mass, respectively. $p^\mu$ and $A^\mu$ are the 4-momentum and 4-vector potential, respectively. We  anticipate that the  Dirac field in the biquaternionic Dirac equation to interact with photons in the  way that Dirac equation couples to photons.

In a biquaternionic form (1) is cast in the form \textcolor[rgb]{0.00,0.07,1.00}{\cite{quaternion,analogy}}
\begin{equation}\label{1}
(\tilde{P}-q\tilde{A}_g)\,\tilde{\gamma} \,\tilde{\Psi}=mc\,\tilde{\Psi}\,,
\end{equation}
where $\tilde{\Psi}$, $\tilde{P}$, $\tilde{A}$ and $\tilde{\gamma}$ are biquaternions defined as
\begin{equation}\label{1}
\tilde{P}=\left(\frac{i}{c}\,E\,, \vec{p}\right)\,,\qquad
\tilde{\gamma}=(i\beta\,, \vec{\gamma})\,,\,\,\,
\tilde{\Psi}=\left(\frac{i}{c}\,\psi_0\,, \vec{\psi}\right)\,,\qquad \tilde{A}_g=\left(\frac{i}{c}\,\varphi_g\,, \vec{A}_g\right),
\end{equation}
where $\vec{A}_g$ and $\varphi_g$ are the vector and scalar potentials of the  boson interacting with the electron. Recall that a biquaternion is generally represented by a scalar and vector components.
In quantum mechanics, the  energy and momentum become operators, \emph{viz.}, $\vec{p}=-i\hbar\vec{\nabla}$ and $E=i\hbar\,\frac{\partial}{\partial t}$. Note that $\vec{\gamma}=\beta\vec{\alpha}$, is a $4\times4$ matrix, where $\beta=\left(
\begin{array}{cc}
1 & 0 \\
0 & -1 \\
\end{array}
\right)$, $\vec{\gamma}=\left(
\begin{array}{cc}
0 & \vec{\sigma} \\
-\vec{\sigma} & 0 \\
\end{array}
\right)$, and $\vec{\sigma}$ are the $2\times2$ Pauli matrices.

In a recent paper \textcolor[rgb]{0.00,0.07,1.00}{\cite{analogy}} we have shown that the biquaternionic Dirac equation produces Maxwell-like equations with electric-like and magnetic-like matter fields that are defined from the linear combinations of the biquaternionic matter fields ($\tilde{\Psi}$), $\psi_0$ and $\vec{\psi}$ and the $\vec{\gamma}$ matrices. The biquaternionic free Dirac equation is obtained from (2) by setting $\vec{A}=0$, \emph{viz.}, $\tilde{P}\,\tilde{\gamma} \,\tilde{\Psi}=mc\,\tilde{\Psi}$. While in the ordinary Dirac equation $\tilde{\Psi}$ are spinor which in 4-dimension has 4-components, $\tilde{\Psi}$ in our present case is a biquaternion consisting of  a scalar and 3-vector totalling to 4-components.

The biquaternion multiplication rule for two biquaternions, $\tilde{A}=(a_0\,, \vec{a})$ and
$\tilde{B}=(b_0\,, \vec{b})$, is given by \textcolor[rgb]{0.00,0.07,1.00}{\cite{quaternion,analogy}}
\begin{equation}\label{1}
\tilde{A}\,\tilde{B}=(a_0\,b_0-\vec{a}\cdot\vec{b}\,\,,
\,\,a_0\,\vec{b}+\vec{a}\,b_0+\vec{a}\times\vec{b})\,,
\end{equation}
 where in general the scalar and vector parts of each biquaternions are complex.

Now apply  (3) and (4) in   (2) and then equate the real and
imaginary parts into the two sides to each other, to obtain
\begin{equation}\label{1}
\hspace{-1cm}\vec{\nabla}\cdot\vec{E}_D=\frac{\rho_D}{\varepsilon_0}-\frac{qc}{\hbar}\vec{A}_g\cdot\vec{B}_D\,,
\end{equation}
\begin{equation}\label{1}
\hspace{-1cm}\vec{\nabla}\times\vec{E}_D+\frac{\partial\vec{B}_D}{\partial
t}=-\frac{q}{\hbar c}\varphi_g\,\vec{E}_D-\frac{qc}{\hbar}\vec{A}_g\,\times\vec{B}_D\,,
\end{equation}
\begin{equation}\label{1}
\hspace{-1cm}\vec{\nabla}\times\vec{B}_D-\frac{1}{c^2}\frac{\partial\vec{E}_D}{\partial
t}\,=\mu_0\vec{J}_D+\frac{q}{\hbar c}\vec{A}_g\times\vec{E}_D-\frac{q}{\hbar c}\,\varphi_g\,\vec{B}_D,
\end{equation}
and
\begin{equation}\label{1}
\vec{\nabla}\cdot\vec{B}_D=\rho_{mD}\,,
\end{equation}
where we denote
\begin{equation}\label{1}
\vec{E}_D=c^2\vec{\gamma}\times\vec{\psi}\,,\,\,\,
\vec{B}_D=\vec{\gamma}\,\psi_0+c\beta\,\vec{\psi}\,,\,\,\,\, \vec{J}_D=\frac{m\,c^2}{\mu_0\hbar}\,\,\vec{\psi}\,,\,\,\,\,
\rho_D=\frac{m}{\mu_0\hbar}\,\psi_0\,,\,\,\, \rho_{mD}=\frac{q}{\hbar c}\,\vec{A}_g\cdot\vec{E}_D,
\end{equation}
provided that \footnote{We can relax this condition by allowing $\Lambda=\beta\psi_{0}+\,\vec{\gamma}\cdot\vec{\psi}$. See Appendix B.}
\begin{equation}
\psi_{0}=-\,c\beta\,\vec{\gamma}\cdot\vec{\psi}\,.
\end{equation}
We call here $\vec{E}_D$ and $\vec{B}_D$ the electric  and magnetic matter (inertial) fields, respectively. $\rho_D$ and $\vec{J}_D$ are the matter and current densities, respectively. As in the ordinary electrodynamics, the electromagnetic fields due to a moving charge are perpendicular to each other and the velocity of the charge. This property is also applied to matter inertial fields. This particular plausible  linear combinations of the vector and scalar, reproduces Maxwell's-like equations. This implies that the biquaternionic Dirac equation is very rich. Maxwell's equations are known to describe spin -1 particle (photon) which is represented by a 3 - vector and a scalar, whereas Dirac's equation describes spin - $\frac{1}{2}$ particles  are represented by 4-component spinor. In our biquaternionic Dirac's equation, our particle is described by 4 - components (a scalar and a 3-vector).

Employing this formalism, the biquaternionic momentum eigen-value equation is shown to reproduce, the Schrodinger, Dirac and Klein-Gordon equations, where the wavefunction and the momentum operator are  generalized to become  biquaternions \textcolor[rgb]{0.00,0.00,1.00}{\cite{uqe}}.
A different view of  a biquaternionic quantum mechanics  with states defined on a biquaternionic Hilbert space has been developed by  Finkelstein \emph{et al}., Adler and  Horwitz \textcolor[rgb]{0.00,0.07,1.00}{\cite{adler,adler1,adler2}}. It was shown in these papers that the biquaternionic quantum mechanics is richer theory than the  ordinary complex quantum mechanics. However, in our present formulation of quantum mechanics, we express our initial equation as an eigen-value quantum   equation but that equation is then reduced to Maxwell-like equations, where we deal with deterministic fields expressed by $\vec{E}_D$ and $\vec{B}_D$ rather than probabilistic wavefunction that the ordinary quantum mechanics works with. The solution of  our biquaternionic Dirac equation reduces to solving the Maxwell-like equations.

It is worth mentioning that when $\vec{A}_g$ and $\varphi_g$ are absent, (5) - (8) reduce to the ordinary  Maxwell's-like equations \textcolor[rgb]{0.00,0.07,1.00}{\cite{analogy}}.
\begin{equation}\label{1}
\hspace{-1cm}\vec{\nabla}\cdot\vec{E}_D=\frac{\rho_D}{\varepsilon_0}\,,
\end{equation}
\begin{equation}\label{1}
\hspace{-1cm}\vec{\nabla}\times\vec{E}_D+\frac{\partial\vec{B}_D}{\partial
t}=0\,,
\end{equation}
\begin{equation}\label{1}
\hspace{-1cm}\vec{\nabla}\times\vec{B}_D-\frac{1}{c^2}\frac{\partial\vec{E}_D}{\partial
t}\,=\mu_0\vec{J}_D,
\end{equation}
and
\begin{equation}\label{1}
\vec{\nabla}\cdot\vec{B}_D=0\,.
\end{equation}
By employing (9), we see that $\vec{E}_D\cdot\vec{B}_D=0$ and $\vec{J}_D\cdot\vec{E}_D=0$, so that the energy conservation equation of the system described by  (11) - (14) is
\begin{equation}
\frac{\partial u_D}{\partial t}+\vec{\nabla}\cdot\vec{S}_D=0\,, \qquad\qquad u_D=\frac{1}{2}\,\varepsilon_0E^2_D+\frac{B^2_D}{2\mu_0}\,,\qquad \vec{S}_D=\frac{\vec{E}_D\times\vec{B}_D}{\mu_0}\,.
\end{equation}
It is interesting that (15) is independent of $\varphi_g$ and $\vec{A}_g$. This implies that these two fields are not dynamical. We see from (9) that magnetic charge (matter) results from interaction of matter fields with the vector potential.

Now take the divergence of (6) and use (5) and (8) to obtain
\begin{equation}
\vec{B}_g\cdot\vec{B}_D-\frac{1}{c^2}\vec{E}_g\cdot\vec{E}_D=\mu_0(\vec{A}_g\cdot\vec{J}_D-\varphi_g\,\rho_D)\,,
\end{equation}
where $\vec{E}_g=-\vec{\nabla}\varphi_g-\frac{\partial \vec{A}_g}{\partial t}$ and $\vec{B}_g=\vec{\nabla}\times\vec{A}_g$.
Taking the divergence of (7) and using (5) and (8) yield
\begin{equation}
\vec{\nabla}\cdot\vec{J}_D+\frac{\partial\rho_D}{\partial t}=-\frac{qc\varepsilon_0}{\hbar }(\vec{E}_g\cdot\vec{B}_D+\vec{B}_g\cdot\vec{E}_D)\,.
\end{equation}

Equations (16) and (17) show that the electromagnetic fields are coupled to the matter inertial and magnetic fields. The matter current and charge densities are coupled to the  vector and scalar potentials, respectively. The continuity equation (mass conservation) dictates that (17) to split into
\begin{equation}
\vec{\nabla}\cdot\vec{J}_D+\frac{\partial\rho_D}{\partial t}=0\,,\qquad\qquad \vec{E}_g\cdot\vec{B}_D+\vec{B}_g\cdot\vec{E}_D=0\,.
\end{equation}
In particular, this equation is satisfied if we take the property that $\vec{E}_g$ is perpendicular to $\vec{B}_D$ and  $\vec{B}_g$ is perpendicular to $\vec{E}_D$.

\section{\textcolor[rgb]{0.00,0.07,1.00}{Symmetric Dirac equation with magnetic matter density}}

In electromagnetism, one can extend Maxwell's equations to incorporate magnetic monopoles (magnetic charges) by writing Maxwell's equations in a symmetric form \textcolor[rgb]{0.00,0.07,1.00}{\cite{griff}}. This form preserves the duality transformation of the electric and magnetic fields, \emph{viz.}, $\vec{E}\rightarrow c\vec{B}$ and $c\vec{B}\rightarrow -\vec{E}$. In the same fashion, we can write (5) - (8) in a symmetric form as
\begin{equation}\label{1}
\vec{\nabla}\cdot\vec{E}_D=\frac{\rho\,'_D}{\varepsilon_0}\,,
\end{equation}
\begin{equation}\label{1}
\vec{\nabla}\times\vec{E}_D=-\frac{\partial \vec{B}_D}{\partial
t}-\vec{J}_{mD}\,,
\end{equation}
\begin{equation}
\vec{\nabla}\times\vec{B}_D=\mu_0\vec{J}\,'_D+\frac{1}{c^2}\frac{\partial
\vec{E}_D}{\partial t}\,,
\end{equation}
and
\begin{equation}
\vec{\nabla}\cdot\vec{B}_D=\rho_{mD}\,,
\end{equation}
where
\begin{equation}
\rho\,'_D=\rho_D-\frac{qc{\varepsilon_0}}{\hbar }\vec{A}_g\cdot\vec{B_D}\,,
\end{equation}
\begin{equation}
\vec{J}_{mD}=\frac{q}{\hbar c}(\varphi_g \,\vec{E}_D+c^2\vec{A}_g\times\vec{B}_D)\,,
\end{equation}
\begin{equation}
\vec{J}\,'_D=\vec{J}_D-\frac{qc\varepsilon_0}{\hbar}(\varphi_g\, \vec{B}_D-\vec{A}_g\times\vec{E}_D)\,.
\end{equation}
We define the quantity $\rho_{mD}$ as the magnetic matter density. This has to be properly interpreted. It is interesting to observe that the sources of the matter fields are of quantum nature. The solution of our Dirac biquaternionic equation is thus a solution of the system of the Maxwell-like equations, (19) - (22). After solving for $\vec{E}_D$ and $\vec{B}_D$ we can then solve for $\psi_0$ and $\vec{\psi}$ using (9) and (10).

The divergence of (24) together with (19), (18) and (22) yield
\begin{equation}
\vec{\nabla}\cdot\vec{J}_{mD}+\frac{\partial\rho_{mD}}{\partial t}=0\,.
\end{equation}
Equations (18) and (26) reveal that inertial and magnetic masses in the presence of interaction are separately conserved.

We conjecture here a possible existence of an electromagnetic system (\emph{e.g}., self-interacting photons) that fulfils  (5) - (8). For such a system one has
\begin{equation}\label{1}
\vec{\nabla}\cdot\vec{E}_g=\frac{\rho}{\varepsilon_0}-\frac{qc}{\hbar}\vec{A}_g\cdot\vec{B}_g\,,
\end{equation}
\begin{equation}\label{1}
\vec{\nabla}\times\vec{E}_g+\frac{\partial \vec{B}_g}{\partial
t}=-\frac{q}{\hbar c}\left(\,\varphi_g \,\vec{E}+c^2\vec{A}_g\times\vec{B}\right)\,,
\end{equation}
\begin{equation}
\vec{\nabla}\times\vec{B}_g-\frac{1}{c^2}\frac{\partial
\vec{E}_g}{\partial t}=\mu_0\vec{J}-\frac{q}{\hbar c}\left(\,\varphi_g\, \vec{B}-\vec{A}_g\times\vec{E}_g\right)\,,
\end{equation}
and
\begin{equation}
\vec{\nabla}\cdot\vec{B}_g=\frac{q}{\hbar c}\vec{A}_g\cdot\vec{E}_g\,.
\end{equation}
Notice here $\vec{J}_{pm}=\frac{q}{\hbar c}(\varphi_g \,\vec{E}_g+c^2\vec{A}_g\times\vec{B}_g)$ can be called the photon magnetic current, since it depends on the vector and scalar potential of the photon.  This is so because the photon behaves as a magnetic charge (monopole). However, an electric photon electric current can also be defined as $\vec{J}_{pe}=\frac{q\varepsilon_0c}{\hbar}\,(\vec{A}_g\times\vec{E}_g-\varphi_g\, \vec{B}_g)$. This is because the photon has an effective charge following its mass. This dual behavior is so because the photon is a particle with mass and charge. The photon once behaves like a magnetic charge ($q_m$) and then like an electric charge ($q_e$), but not both simultaneously. These two dual behaviors are related by the uncertainty relation that $q_mq_e=\hbar/2$ that is coined in the Dirac's quantization rule \textcolor[rgb]{0.00,0.07,1.00}{\cite{dirac}}. It is obvious that the ordinary electrodynamics is restored when $\vec{A}_g$ and $\varphi_g=0$. It is interesting that the two currents, $\vec{J}_{pm}$ and $\vec{J}_{pe}$ are non-dissipative currents since they do not appear in the energy conservation equation of the system of equations, Eqs.(27) - (30).

\section{\textcolor[rgb]{0.00,0.07,1.00}{Continuity equation and static fields}}
 It is interesting to see that (27) - (30) can be seen as Maxwell's equations in a background identified by $\vec{A}_g$ and $\varphi_g$.  Equations similar to (28) and (29) are obtained by Tiwari \textcolor[rgb]{0.00,0.07,1.00}{\cite{tiwari}}.

For a pure electromagnetic system (we omit the subscripts $g$ and $D$ and $m$ will be the mass of the boson field), the continuity equation in (21) will be transformed into
\begin{equation}
\vec{\nabla}\cdot\vec{J}+\frac{\partial\rho}{\partial t}=-\frac{2q}{\hbar\mu_0c}\,\vec{E}\cdot\vec{B}\,.
\end{equation}
Hence, charge conservation is violated whenever the term, $\vec{E}\cdot\vec{B}\ne0\,$. The presence of this term  is found to be associated with the CP (Charge conjugation - Parity) symmetry violation in quantum-chromodynamics \textcolor[rgb]{0.00,0.07,1.00}{\cite{pecci}}. It is interesting to see the quantum nature of this violating term. In free space $\rho=0$ and $\vec{J}=0$, so that $\vec{E}\cdot\vec{B}=0$. It is interesting to note that the electrodynamics in (27) - (30) is  invariant under parity and time-reversal transformations. It is also invariant under CP transformation. The above system described by  (27) - (30) is Lorentz invariant, and gauge invariant, where $\vec{A}'=\vec{A}-\vec{\nabla}\Lambda,\,\,\, \varphi'=\varphi+\frac{\partial\Lambda}{\partial t}$, provided that \begin{equation}\dot\Lambda\,\vec{B}=-\vec{\nabla}\Lambda\times\vec{E}\,,\qquad\qquad\,\dot\Lambda\,\vec{E}=c^2\vec{\nabla}\Lambda\times\vec{B}\,,\qquad \vec{\nabla}\Lambda\cdot\vec{E}=0\,,\qquad \vec{\nabla}\Lambda\cdot\vec{B}=0\,.
\end{equation}
They can be expressed in a covariant form that guarantees Lorentz invariance. Note that the electric and magnetic fields due to a moving charge with velocity $\vec{v}$ are given by $\vec{E}=\vec{v}\times\vec{B}$ and $\vec{B}=-\frac{\vec{v}\times\vec{E}}{c^2}$. Moreover, these fields are perpendicular to the direction of motion. Hence, (32) suggests that $\Lambda$ should satisfy the relation
$$\vec{v}=\frac{c^2\vec{\nabla}\Lambda}{\dot \Lambda}\,.$$

It is interesting to note that static fields electrodynamics is described by
\begin{equation}\label{1}
\vec{\nabla}\cdot\vec{E}=\frac{\rho}{\varepsilon_0}-\frac{qc}{\hbar}\vec{A}\cdot\vec{B}\,,
\end{equation}
\begin{equation}\label{1}
\vec{\nabla}\times\vec{E}=-\frac{q}{\hbar c}\left(\,\varphi \,\vec{E}+c^2\vec{A}\times\vec{B}\right)\,,
\end{equation}
\begin{equation}
\vec{\nabla}\times\vec{B}=\mu_0\vec{J}-\frac{q}{\hbar c}\left(\,\varphi\, \vec{B}-\vec{A}\times\vec{E}\right)\,,
\end{equation}
and
\begin{equation}
\vec{\nabla}\cdot\vec{B}=\frac{q}{\hbar c}\vec{A}\cdot\vec{E}\,.
\end{equation}
The electromagnetic currents (the two bracket in the right-hand side in (34) and (35)), are non-dissipative. Notice that in Chert-Simon theory these two currents are of topological origin \textcolor[rgb]{0.00,0.07,1.00}{\cite{chern2}}.

\subsection{\textcolor[rgb]{0.50,0.00,0.50}{Duality transformation}}

Interestingly  in free space, $\vec{J}=0$ and $\rho=0$, there exists non-zero effective charge and current densities arising from the interactions. These are $\rho_{pm}=\frac{q}{\hbar c}\vec{A}\cdot\vec{E}$, $\rho_{pe}=-\frac{qc\varepsilon_0}{\hbar}\vec{A}\cdot\vec{B}$,  $\vec{J}_{pe}$ and $\vec{J}_{pm}$, respectively. In this case  (27) - (30) look like symmetrized Maxwell's equations. They are invariant under duality transformation, $\vec{E}\rightarrow c\vec{B}$\,, $c\vec{B}\rightarrow -\vec{E}$\,, $\varphi\rightarrow\varphi$\,, $\vec{A}\rightarrow\vec{A}$\,, $\vec{J}_{pe}\rightarrow \varepsilon_0c\vec{J}_{pm}$\,, and $\vec{J}_{pm}\rightarrow-\frac{1}{c\varepsilon_0}\,\vec{J}_{pe}$. The application of duality transformation in axion electrodynamic is studied recently by Visinelli \textcolor[rgb]{0.00,0.07,1.00}{\cite{axion}}.

\section{\textcolor[rgb]{0.00,0.07,1.00}{Axion electrodynamics}}
Wilczek has studied the electrodynamics of axions by adding a Lagrangian term to Maxwell's lagrangian of the form ${\cal L}=\kappa\,\theta\vec{E}\cdot\vec{B}$, where $\theta$ is the axion field and $\kappa$ is a dimensionless constant. He obtained the following equations \textcolor[rgb]{0.00,0.07,1.00}{\cite{chern1}}
\begin{equation}
\vec{\nabla}\cdot\vec{E}=\frac{\rho}{\varepsilon_0}-c\kappa\vec{\nabla}\theta\cdot\vec{B}\,,
\end{equation}
\begin{equation}
\vec{\nabla}\times\vec{E}+\frac{\partial \vec{B}}{\partial t}=0\,,
\end{equation}
\begin{equation}
\vec{\nabla}\times\vec{B}-\frac{1}{c^2}\frac{\partial \vec{E}}{\partial t}=\mu_0\vec{J}+\frac{\kappa}{c}\,(\dot \theta\,\vec{B}+\vec{\nabla}\theta\times\vec{E})\,,
\end{equation}
and
\begin{equation}
\vec{\nabla}\cdot\vec{B}=0\,,
\end{equation}
where  $\dot\theta=\frac{\partial\theta}{\partial t}\,$. The importance of axions is noted in the field of cosmology, CD and condensed matter physics.

Interestingly, if we compare (37) - (40) with (27) - (30) then
\begin{equation}
\dot \theta=-\frac{q}{\hbar\kappa }\,\varphi\,,\qquad \vec{\nabla} \theta=\frac{q}{\hbar\kappa  }\,\vec{A}\,,\qquad\varphi\,\vec{E}=-c^2\vec{A}\times\vec{B}\,,\qquad \vec{A}\cdot\vec{E}=0\,.
\end{equation}
Hence, axion fields coupled to electromagnetic field in an analogous manner  photons do. (41) reveals that $\dot\theta$ represents some energy scale of some system, and $\vec{\nabla}\theta$ represents its momentum. It is shown by Li {\it et al.} that magnetic fluctuations of topological insulators couple to the electromagnetic fields exactly like the axions \textcolor[rgb]{0.00,0.07,1.00}{\cite{insulator}}. It is interesting to see that the phase difference
$$
\Delta\theta=\int\vec{\nabla}\theta\cdot d\vec{\ell}=\frac{q}{\kappa\hbar}\,\int\vec{A}\cdot d\vec{\ell}=\frac{q}{\kappa\hbar}\,\phi_B\,.
$$
If we assume that $\phi_B=\frac{h}{q}\,n$, then $\Delta\theta=\frac{2\pi}{\kappa}\,n$\,. This may indicate that $n/\kappa$ is an integer. The two potentials, $\vec{A}$ and $\varphi$ could be different from those ones defined in electromagnetism.

Equation (41) implies that
\begin{equation}
\vec{E}=\frac{(\vec{\nabla}\theta)}{\dot \theta} \,c^2\,\times\vec{B}\,.
\end{equation}
This means that the electric field is perpendicular to the  $\vec\nabla \theta$ and $\vec{B}$. Since in a moving frame the magnetic field is perpendicular to the velocity vector by the relation $\vec{E}=\vec{v}\times\vec{B}$, then $\vec{v}=\frac{(\vec{\nabla}\theta)}{\dot \theta} \,c^2$\,. This relation can be compared with the relativistic relation, $\vec{v}=\frac{\vec{p}}{E}\,c^2$, where $E$ is the total relativistic energy of the particle. This relation implies that, $\vec{v}\propto \vec{\nabla}\theta$, \emph{i.e.}, the gradient of $\theta$ points along the velocity direction of the axion field. The velocity is positive for a growing scalar axion field in space and time simultaneously. It is negative otherwise.
It is interesting to see that (42) is compatible with (32) if we let $\theta=\Lambda$. Hence, (32) further implies that $\vec{E}=\vec{v}\times\vec{B}$ and $\vec{B}=-\frac{\vec{v}}{c^2}\times\vec{E}$.

It is remarkable to see that the axion electrodynamics can be obtained from (33) - (36) by gauging them and allowing  $\Lambda=\theta$\,, $\vec{A}=0$ and $\varphi=0$\, such that $\vec{\nabla}\Lambda\cdot\vec{E}=0$ and $\dot\Lambda\,\vec{E}=c^2\vec{\nabla}\Lambda\times\vec{B}$. This is consistent with (41) and (42).

Applying (41) in  (27) - (30) yields  a symmetric Wilczek axion electrodynamics as
\begin{equation}
\vec{\nabla}\cdot\vec{E}=\frac{\rho}{\varepsilon_0}-c\kappa\vec{\nabla}\theta\cdot\vec{B}\,,
\end{equation}
\begin{equation}
\vec{\nabla}\times\vec{E}=-\frac{\partial \vec{B}}{\partial t}-\frac{\kappa}{c}\left(-\dot\theta\vec{E}+c^2\vec{\nabla}\theta\times\vec{B}\right)\,,
\end{equation}
\begin{equation}
\vec{\nabla}\times\vec{B}=\frac{1}{c^2}\frac{\partial \vec{E}}{\partial t}+\mu_0\vec{J}+\frac{\kappa}{c}\,\left(\dot \theta\,\vec{B}+\vec{\nabla}\theta\times\vec{E}\right)\,,
\end{equation}
and
\begin{equation}
\vec{\nabla}\cdot\vec{B}=\frac{\kappa}{c}\,\vec{\nabla}\theta\cdot\vec{E}\,,
\end{equation}
Differentiating the first equation in (41) and taking the divergence of the second equation in (41) yield
\begin{equation}
\frac{1}{c^2}\frac{\partial^2\theta}{\partial t^2}-\nabla^2\theta+\frac{q}{\hbar\kappa}\left(\vec{\nabla}\cdot\vec{A}+\frac{1}{c^2}\frac{\partial\varphi}{\partial t}\right)=0\,.
\end{equation}
If axions satisfy the Klein-Gordon equation then the Lorenz gauge condition  will be modified to
\begin{equation}
\vec{\nabla}\cdot\vec{A}+\frac{1}{c^2}\frac{\partial\varphi}{\partial t}=\kappa\left(\frac{m^2c^2}{q\hbar}\right)\theta\,.
\end{equation}
Note that the expression $B_c=\frac{m^2c^2}{q\hbar}$ is known as the Schwinger critical field \textcolor[rgb]{0.00,0.07,1.00}{\cite{schwinger}}.  Therefore, the interaction of axions gives rise to a residual magnetic field, $B_c$. Thus, if we know the field the axions create, we can estimate their masses, or vice versa. We see from (48) that the axions field is coupled to the electromagnetic field (photon). We have recently shown that the violation of Lorenz gauge condition leads to interesting consequences \textcolor[rgb]{0.00,0.07,1.00}{\cite{lorenz}}. Thus, if axions are massless then the Lorenz gauge condition is satisfied. Therefore, the relaxation of the  Lorenz gauge condition can be used to study massive axions. In such a case axions act like massive photon in electrodynamics making the electromagnetic range finite. One can therefore relate the vector and scalar potentials, $\vec{A}$ and $\varphi$ to the axion massive fields. Moreover, axions are coupled to the electromagnetic field as is evident from (33) - (36).

In superconductivity the electromagnetic force is of short range. Thus, if axions exist inside superconductors, they will give rise to effects similar to those induced by massive photons. Hence, the residual magnetic field developed by axions, $B_c$, could be the critical magnetic field observed in superconductors. Moreover, if axions are present today, then the microwave background radiation temperature can set a limit on their masses. The energy density due to axions is  $u=B_c^2/(2\mu_0)$. This leads to a limit that $m<4.077\times 10^{-38}\rm kg$ ( $\rm 0.023\, eV/c^2 )$\,.

It is interesting to note that despite the presence of interaction, the energy conservation equation of the system in (32) - (35) is the same as that of the ordinary electrodynamics. This implies that axions are not dynamical fields. However, the energy conservation equation of Wilczek equations, (43) - (46), involves  spatial and temporal variations of the axions field $\theta$.

Let us now rewrite (43) -(46) as \footnote{See Appendix A for alternative expression}
\begin{equation}
\vec{\nabla}\cdot\vec{\cal E}=\frac{\tilde{\rho}}{\varepsilon_0}\,,
\end{equation}
\begin{equation}
\vec{\nabla}\times\vec{\cal E}=-\frac{\partial \vec{\cal B}}{\partial t}-\vec{\cal J}_m\,,
\end{equation}
\begin{equation}
\vec{\nabla}\times\vec{\cal B}-\frac{1}{c^2}\frac{\partial \vec{\cal E}}{\partial t}=\mu_0\vec{\tilde{J}}\,,
\end{equation}
and
\begin{equation}
\vec{\nabla}\cdot\vec{\cal B}=\tilde{\rho}_m\,,
\end{equation}
where
\begin{equation}
\vec{\cal E}=\vec{E}+\kappa c\,\theta\,\vec{B}\,\,,\qquad\qquad \vec{\cal B}=\vec{B}-\frac{\kappa}{c}\,\theta\vec{E}\,,
\end{equation}
and
\begin{equation}
\hspace{0cm}\vec{\tilde{J}}=\vec{J}+\frac{\kappa\,\theta}{\mu_0c}\,\vec{J}_m\,, \tilde{\rho} =\rho+\frac{\kappa}{\mu_0c}\,\rho_m+\frac{\kappa}{\mu_0c}\,\vec{\nabla}\theta\cdot\vec{B}\,,\,\, \tilde{\rho}_m= -\frac{\kappa\,\theta}{\varepsilon_0c}\rho-\frac{\kappa}{c}\vec{\nabla}\theta\cdot\vec{E}\,, \vec{\cal J}_m=\frac{\kappa}{c}\,(-\dot\theta\vec{E}+c^2\vec{\nabla}\theta\times\vec{B})\,.
\end{equation}
Notice that $\vec{{\cal E}}$ and $\vec{{\cal B}}$ are connected by duality transformation when $\vec{E}$ and $\vec{B}$ are dually transformed. The above equations reduce to the ordinary Maxwell's equations when $\theta=0$. However, if $\theta=\rm const.$, then $\vec{\tilde{J}}=\vec{J}\,,\,\,\, \tilde{\rho} =\rho(1-\kappa^2\theta)\,,\,\,\, \tilde{\rho}_m= -\frac{\kappa\,\theta}{\varepsilon_0c}\rho\,,\,\,\, \vec{\cal J}_m=0\,.$ Hence, (49) - (52) become
$$
\vec{\nabla}\cdot\vec{\cal E}=\frac{(1-\kappa^2\theta)\,\rho}{\varepsilon_0}\,,\qquad
\vec{\nabla}\times\vec{\cal E}=-\frac{\partial \vec{\cal B}}{\partial t}\,,
$$
and
$$
\vec{\nabla}\times\vec{\cal B}-\frac{1}{c^2}\frac{\partial \vec{\cal E}}{\partial t}=\mu_0\vec{J}\,,
\qquad
\vec{\nabla}\cdot\vec{\cal B}=-\frac{\kappa\,\theta}{\varepsilon_0c}\,\rho\,.
$$
Charge conservation in (31) can be restored if we expressed (31) and (48) in the form
\begin{equation}
\vec{\nabla}\cdot\vec{J}_T+\frac{\partial\rho_T}{\partial t }=0\,,
\end{equation}
where
\begin{equation}
\vec{J}_T=\vec{J}-\alpha\, \vec{A}\,,\qquad \rho_T=\rho-\frac{\alpha}{c^2}\,\varphi\,,
\end{equation}
and hence
\begin{equation}
\theta=-\frac{2q^2\varepsilon_0}{\alpha\kappa m^2c}\,\vec{E}\cdot\vec{B}\,,
\end{equation}
where $\alpha$ is a constant. It is thus the total current $\vec{J}_T$ that is conserved.
Notice that in free space (31) implies that charge conservation is restored. In a medium filled with axion field, the electric and magnetic fields are no longer transverse as those in free space. Equation (57) also shows that the source of the axions field is electromagnetic. Moreover, the axions field interacts equally with negative and positive charges. In London's theory of superconductivity, $\alpha=\frac{nq^2}{m}$, where $n$ is a number density, and hence (57) can be written as
\begin{equation}
\theta=-\frac{2\,\varepsilon_0}{\kappa nmc}\,\vec{E}\cdot\vec{B}\,,
\end{equation}
Apparently, $\theta$ is odd under parity and time-reversal transformation. It is even under TP transformation. It is also found to be responsible for  the absence of CP violation symmetry in CD \textcolor[rgb]{0.00,0.07,1.00}{\cite{pecci}}. Equations (58) shows that axions  do not occur in  empty space where $\vec{E}\cdot\vec{B}=0$, and that the Lorenz gauge condition is restored, as evident from (48).

\subsection{\textcolor[rgb]{0.50,0.00,0.50}{Pure magnetic system}}

Let us consider now a system in which the electric field vanishes, i.e., $\vec{E}=0$ and $\varphi=mc^2/q$. Equation (31) shows that the electric charge is conserved.
Substituting this in (27) - (30) yields
\begin{equation}\label{1}
\rho=\frac{qc \,\varepsilon_0}{\hbar}\,\vec{A}\cdot\vec{B}\,,
\end{equation}
\begin{equation}\label{1}
\frac{\partial \vec{B}}{\partial
t}=-\frac{qc}{\hbar}\,\vec{A}\times\vec{B}\,,
\end{equation}
\begin{equation}
\vec{\nabla}\times\vec{B}=\mu_0\vec{J}-\frac{mc}{\hbar }\, \vec{B}\,,
\end{equation}
and
\begin{equation}
\vec{\nabla}\cdot\vec{B}=0.
\end{equation}
One can associate an angular velocity of (60) for the precession of the magnetic field given by $\vec{\omega}=\frac{qc}{\hbar}\vec{A}$\,.  Hence, an effective charge density in (59) yields $\rho=\varepsilon_0\,\vec{\omega}\cdot\vec{B}$\,.

The solution of the above equations yields
\begin{equation}\label{1}
\nabla^2\vec{B}+\left(\frac{mc}{\hbar}\right)^2\vec{B}=\mu_0\left(\frac{mc}{\hbar}\vec{J}-\vec{\nabla}\times\vec{J}\right)\,.
\end{equation}
 This pattern of solution is also obtained by Carroll  \emph{et al.} \textcolor[rgb]{0.00,0.07,1.00}{\cite{etal}} for static fields  arising from stationary neutral source, $\rho=0$, with $\vec{\nabla}\cdot\vec{J}=0$. Now if, $\vec{\nabla}\times\vec{J}=\frac{mc}{\hbar}\,\vec{J}$\,, then $\vec{B}$ and $\vec{J}$ are sinusoidal ($\propto\sin (\vec{k}\cdot\vec{r}+\phi))$ with $k=mc/\hbar$ and $\phi$=constant.

\subsection{\textcolor[rgb]{0.50,0.00,0.50}{Pure electric system}}
Consider a system in which $\vec{B}=0$ and that $\varphi=mc^2/q$. Equation (31) shows that the electric charge is conserved.
In this case (27) - (30) yield
\begin{equation}\label{1}
\vec{\nabla}\cdot\vec{E}=\frac{\rho}{\varepsilon_0}\,,
\end{equation}
\begin{equation}\label{1}
\vec{\nabla}\times\vec{E}=-\frac{mc}{\hbar} \,\vec{E}\,,
\end{equation}
\begin{equation}
\frac{\partial
\vec{E}}{\partial t}=-\frac{\vec{J}}{\varepsilon_0}-\frac{qc}{\hbar }\vec{A}\times\vec{E}\,,
\end{equation}
and
\begin{equation}
\vec{A}\cdot\vec{E}=0\,.
\end{equation}
Consistency of (64) and (65) requires that $\rho=0$\,, and hence (31) yields $\vec{\nabla}\cdot\vec{J}=0$, consequently one finds
\begin{equation}\label{1}
\nabla^2\vec{E}+\left(\frac{mc}{\hbar}\right)^2\vec{E}=0\,.
\end{equation}
Now if the electric field is static, then the energy conservation equation (or (66)) yields the relation $\vec{J}\cdot\vec{E}=0$\,. Equations (63) and (68) are noted by \textcolor[rgb]{0.00,0.07,1.00}{\cite{etal}} to arise in magnetohydrodynamics. Hence, the electric and magnetic fields in (63), with $\vec{\nabla}\times\vec{J}=\frac{mc}{\hbar}\,\vec{J}$\,, and (68) satisfy the Helmholtz equation.

\section{\textcolor[rgb]{0.00,0.07,1.00}{Maxwell's equations in moving reference frame}}

Let us now consider that $\vec{p}\,'=\vec{p}-q\vec{A}=0$, or $q\vec{A}=m\vec{v}$, in (27) - (30). This yields
\begin{equation}\label{1}
\vec{\nabla}\cdot\vec{E}=\frac{\rho}{\varepsilon_0}-\frac{mc}{\hbar}\,\vec{v}\cdot\vec{B}\,,
\end{equation}
\begin{equation}\label{1}
\vec{\nabla}\times\vec{E}=-\frac{\partial \vec{B}}{\partial
t}-\frac{mc}{\hbar}\left(\vec{E}+\vec{v}\times\vec{B}\right)\,,
\end{equation}
\begin{equation}
\vec{\nabla}\times\vec{B}=\mu_0\vec{J}+\frac{1}{c^2}\frac{\partial
\vec{E}}{\partial t}-\frac{mc}{\hbar}\left(\vec{B}-\frac{\vec{v}}{c^2}\,\times\vec{E}\right)\,,
\end{equation}
and
\begin{equation}
\vec{\nabla}\cdot\vec{B}=\frac{mc}{\hbar}\,\vec{v}\cdot\vec{E}\,.
\end{equation}
Recall that in a moving frame with respect to the charge,  the electric and magnetic fields  are defined as $\vec{E}\,'=\vec{E}+\vec{v}\times\vec{B}$ and $\vec{B}\,'=\vec{B}-\frac{\vec{v}}{c^2}\,\times\vec{E}\,$ \textcolor[rgb]{0.00,0.07,1.00}{\cite{griff}}. Moreover, $\vec{v}\cdot\vec{E}=0$ and $\vec{v}\cdot\vec{B}=0$, $\vec{E}\,'=0$ and $\vec{B}\,'=0$ for transverse field created by a moving charge. In this case (69) - (72) reduce to the ordinary Maxwell's equations. The same occurs for massless photon ($m=0$). Hence, (69) - (72) represent Maxwell's equations  including the contribution of the massive photons. The latter fields appeared as quantum corrections to the ordinary Maxwell's equations. The correction terms  vanish for massless photons and accordingly  the ordinary Maxwell's equations are restored.

Equations (70) and (71) suggest two kinds of currents. These are electric and magnetic currents. They can be defined as follows:
$$\hspace {3cm} \vec{J}_m=\frac{mc}{\hbar}\left(\vec{E}+\vec{v}\times\vec{B}\right)\,,\qquad\qquad \vec{J}_e=\frac{mc}{\mu_0\hbar}\left(-\vec{B}+\frac{\vec{v}}{c^2}\times\vec{E}\right). \hspace{0.8cm} (A)$$
They are independent of the  charge of the moving particle but depend on its mass. They thus manifest the effect of the  quantum inertial mass on the electrodynamics that is normally ignored. Such currents could lead to kinetic inductance exhibited in some electronic systems \textcolor[rgb]{0.00,0.07,1.00}{\cite{super}}. Therefore, (A) could have interesting consequences when taken into consideration. We notice that magnetic current and charge are present whenever the velocity of the moving charge makes and angle (not right) with the electric and magnetic fields. Equation (A) suggests a Hall-like transverse current given by $\vec{J}_{tr.}=\frac{m}{\hbar c\mu_0}\,\vec{v}\times\vec{E}$. This also suggests a transverse conductivity, $\sigma_{tr.}=\frac{mv}{\hbar c\mu_0}$.

Let us now consider a stationary particle, \emph{i.e.,} the case when $\vec{v}=0$. This makes (69) - (72) reduce to
\begin{equation}\label{1}
\vec{\nabla}\cdot\vec{E}=\frac{\rho}{\varepsilon_0}\,,
\end{equation}
\begin{equation}\label{1}
\vec{\nabla}\times\vec{E}=-\frac{\partial \vec{B}}{\partial
t}-\frac{mc}{\hbar}\,\vec{E}\,,
\end{equation}
\begin{equation}
\vec{\nabla}\times\vec{B}=\mu_0\vec{J}+\frac{1}{c^2}\frac{\partial
\vec{E}}{\partial t}-\frac{mc}{\hbar}\,\vec{B}\,,
\end{equation}
and
\begin{equation}
\vec{\nabla}\cdot\vec{B}=0\,.
\end{equation}
The consistency of the above system reveals that $\rho=0$ and $\vec{\nabla}\cdot\vec{J}=0$. Upon using (36), one finds that $\vec{E}\cdot\vec{B}=0$. The solution of (73) - (76) is
\begin{equation}
\frac{1}{c^2}\frac{\partial^2\vec{E}}{\partial t^2}-\nabla^2\vec{E}-\left(\frac{mc}{\hbar}\right)^2\vec{E}=\frac{\partial
}{\partial t}\left(\,\frac{2mc}{\hbar}\,\vec{B}-\mu_0\vec{J}\,\right)\,.
\end{equation}
Thus, the electric field satisfies Klein-Gordon equation with an imaginary mass (tachyons) provided that
\begin{equation}
\vec{J}=\frac{2mc}{\mu_0\hbar}\,\vec{B}\,.
\end{equation}
This relation is normally reflected in a chiral magnetic effect \textcolor[rgb]{0.00,0.07,1.00}{\cite{cme1}}. In this case one can define a  magnetic  conductivity, $\sigma_c=\frac{2mc}{\mu_0\hbar}$. This can be written as $\sigma_c=\frac{2mc^2}{q\varphi}\,\sigma_m$.

Substituting (78) in (75) to obtain the corresponding Maxwell's equations for massive photon
\begin{equation}\label{1}
\vec{\nabla}\cdot\vec{E}=0\,,
\end{equation}
\begin{equation}\label{1}
\vec{\nabla}\times\vec{E}=-\frac{\partial \vec{B}}{\partial
t}-\frac{mc}{\hbar}\,\vec{E}\,,
\end{equation}
\begin{equation}
\vec{\nabla}\times\vec{B}=\frac{1}{c^2}\frac{\partial
\vec{E}}{\partial t}+\frac{mc}{\hbar}\,\vec{B}\,,
\end{equation}
and
\begin{equation}
\vec{\nabla}\cdot\vec{B}=0\,.
\end{equation}
The above equations reveal that the magnetic field also satisfies Klein-Gordon's equation with an imaginary mass (tachyons). However, static fields in (79) - (82) yield the two equations, $\vec{\nabla}\times\vec{E}=-\frac{mc}{\hbar}\,\vec{E}$ and  $\vec{\nabla}\times\vec{B}=\frac{mc}{\hbar}\,\vec{B}$ that are solved to give $\nabla^2\vec{E}+\left(\frac{mc}{\hbar}\right)^2\vec{E}=0\,$ and $\,\nabla^2\vec{B}+\left(\frac{mc}{\hbar}\right)^2\vec{B}=0$. These are similar to (69) and (72).

Now by defining the electromagnetic field, $\vec{F}=\vec{E}+c\vec{B}\,i$ \textcolor[rgb]{0.00,0.07,1.00}{\cite{vector, vectors}}, (27) -(30), for the vacuum case with $\varphi=mc^2/q$ and $\vec{A}=0$, yield
\begin{equation}\label{1}
\vec{\nabla}\cdot\vec{F}=0\,,
\end{equation}
and
\begin{equation}\label{1}
i\hbar\frac{\partial \vec{F}}{\partial
t}=c\hbar\,\vec{\nabla}\times\vec{F}+mc^2\,\vec{F}\,.
\end{equation}
Equation (84) can be written as
\begin{equation}\label{1}
i\hbar\frac{\partial \vec{F}}{\partial
t}=-ic\hbar\,\vec{S}\cdot\vec{\nabla}\vec{F}+mc^2\,\vec{F}\,,
\end{equation}
where $\vec{S}$ is spin-1 matrices \textcolor[rgb]{0.00,0.07,1.00}{\cite{vector, curl}}. This suggests that (85) is a quantum equation for massive photon whose Hamiltonian is defined by
\begin{equation}\label{1}
H=c\vec{S}\cdot\vec{p}+mc^2\,.
\end{equation}
Moreover, as evident from (85), $\vec{F}$ satisfies the Klein-Gordon equation with a source as
\begin{equation}\label{1}
\frac{1}{c^2}\frac{\partial^2 \vec{F}}{\partial
t^2}-\nabla^2\vec{F}+\left(\frac{mc}{\hbar}\right)^2\vec{F}=-\frac{2mc}{\hbar}\,\vec{\nabla}\times\vec{F}\,.
\end{equation}
Using (84) this becomes
\begin{equation}\label{1}
\frac{1}{c^2}\frac{\partial^2 \vec{F}}{\partial
t^2}-\nabla^2\vec{F}+\frac{2mi}{\hbar}\,\frac{\partial\vec{F}}{\partial t}-\left(\frac{mc}{\hbar}\right)^2\vec{F}=0\,.
\end{equation}
It is interesting that (88) is the Dirac's equation for a free spin-1/2 particle which is normally expressed as a first-order differential equation
\begin{equation}\label{1}
i\hbar\frac{\partial \psi}{\partial
t}=-ic\hbar\,\vec{\gamma}\cdot\vec{\nabla}\psi+\beta mc^2\,\psi\,,
\end{equation}
where $\psi$ is the spinor. By taking the second term in the right hand-side in (89) to the other side, and squaring the two sides of the resulting equation, we obtain an equation of the form in (88). Despite the fact that Dirac's equation describes a spin-1/2  particles and Maxwell's equations describe spin-1 particle which belong to distinct irreducible representations of the Poincare group, they exhibit a similarity shown above. Thus, when $\beta=1$, $\vec{\gamma}\rightarrow\vec{S}$, $\psi\rightarrow \vec{F}$, and $m_e\rightarrow m$. This urges us to explore this deep connection.
Thus, (88) can be seen as  the quantum equation of the massive photon. It is worth to remark that moving the second term in the right - hand side in  (89) to the left side and squaring the resulting operator equation yield \textcolor[rgb]{0.00,0.07,1.00}{\cite{uqe}}.
\begin{equation}\label{1}
\frac{1}{c^2}\frac{\partial^2 \psi}{\partial
t^2}-\nabla^2\psi+\frac{2m\beta i}{\hbar}\,\frac{\partial\psi}{\partial t}-\left(\frac{mc}{\hbar}\right)^2\psi=0\,.
\end{equation}
Because of the $\beta$ term in (90), we have two solutions (particle-antiparticle), whereas (88) assumes only one solution. This is because the anti-photon is the same as the photon, while the anti-electron is a positron.

The energy conservation equation associated with (84) can be written as
\begin{equation}\label{1}
\vec{\nabla}\cdot\left(\frac{\vec{F}\times\vec{F}^*}{\mu_0}\right)+\frac{\partial }{\partial
t}\left(\frac{-i}{\mu_0c}\,\vec{F}\cdot\vec{F}^*\right)=0\,.
\end{equation}
This shows that the energy conservation equation for massive photon in vacuum is the same as that for massless photon.
\subsection{\textcolor[rgb]{0.50,0.00,0.50}{Static fields}}
Consider now static electric and magnetic fields. Hence, (69) - (72) reduce to
\begin{equation}\label{1}
\vec{\nabla}\cdot\vec{E}=\frac{\rho}{\varepsilon_0}-\frac{mc}{q_m\hbar}P_m\,,\qquad\qquad P_m=\,q_m\vec{v}\cdot\vec{B}\,,
\end{equation}
\begin{equation}\label{1}
\hspace{1.5cm}\vec{\nabla}\times\vec{E}=-\frac{mc}{q_e\hbar}\vec{F}_e\,,\qquad\qquad\qquad \vec{F}_e=q_e\left(\vec{E}+\vec{v}\times\vec{B}\right)\,,
\end{equation}
\begin{equation}
\hspace{2cm}\vec{\nabla}\times\vec{B}=\mu_0\vec{J}-\frac{mc}{q_m\hbar}\vec{F}_m\,,\qquad\qquad \vec{F}_m=q_m\left(\vec{B}-\frac{\vec{v}}{c^2}\,\times\vec{E}\right)\,,
\end{equation}
and
\begin{equation}
\vec{\nabla}\cdot\vec{B}=\frac{m}{q_ec\hbar}\,P_e\,,\qquad\qquad P_e=q_e\vec{v}\cdot\vec{E}\,.
\end{equation}
Interestingly, (92) - (95) involve the fields, the force and the power on the moving particle. Taking the dot product of the velocity with  (94) and using (92) yield
\begin{equation}\label{1}
\vec{\nabla}\cdot\left(\vec{E}+\vec{v}\times\vec{B}\right)=\frac{1}{\varepsilon_0}\left(\,\rho-\frac{\vec{v}}{c^2}\cdot\vec{J}\,\right)\,,
\end{equation}
This can be expressed as Gauss's law in the moving frame
\begin{equation}\label{1}
\vec{\nabla}\cdot\vec{E}\,'=\frac{\rho\,'}{\varepsilon_0}\,,\qquad\qquad\qquad\vec{E}\,'=\vec{E}+\vec{v}\times\vec{B}\,,\qquad\qquad \rho\,'=\rho-\frac{\vec{v}}{c^2}\cdot\vec{J}\,.
\end{equation}
Similarly taking the divergence of (94) and using the fact that $\vec{\nabla}\cdot\vec{J}=0$ one obtains
\begin{equation}
\hspace{-4cm}\vec{\nabla}\cdot\vec{B}\,'=0\,,  \qquad \qquad\qquad\vec{B}\,'=\vec{B}-\frac{\vec{v}}{c^2}\times\vec{E}\,.
\end{equation}
Taking the cross product of the velocity with (93) and using (92) and (94) yield Ampere's and Faraday's equations in the moving frame
\begin{equation}
\hspace{-2cm}\vec{\nabla}\times\vec{B}\,'=\mu_0\vec{J}\,'\,,\qquad \qquad\vec{\nabla}\times\vec{E}\,'=0\,,\qquad\qquad \vec{J}\,'=\vec{J}-\rho\,\vec{v}\,.
\end{equation}
where the two vector identities $\vec{\nabla}\times(\vec{v}\times\vec{E})=\vec{v}\,(\vec{\nabla}\cdot\vec{E})-(\vec{v}\cdot\vec{\nabla})\vec{E}$ and $\vec{v}\times(\vec{\nabla}\times(\vec{E})=\vec{\nabla}(\vec{v}\,\cdot\vec{E})-(\vec{v}\cdot\vec{\nabla})\vec{E}$ are employed. Therefore, (92) - (95) are equivalent to Maxwell's equations in a moving frame with constant velocity where the fields are static.
\section{\textcolor[rgb]{0.00,0.07,1.00}{Maxwell's equations inside a medium}}
The system of equations, (27) - (30),  can be seen as Maxwell's equations inside a medium with polarization and magnetization vectors, $\vec{P}$ and $\vec{M}$, respectively. In this case Maxwell's equations  read (omitting the subscript $g$)
\begin{equation}\label{1}
\vec{\nabla}\cdot\vec{E}=\frac{1}{\varepsilon_0}\left(\rho-\vec{\nabla}\cdot\vec{P}\right)\,,
\end{equation}
\begin{equation}\label{1}
\vec{\nabla}\times\vec{E}=-\frac{\partial \vec{B}}{\partial
t}\,,
\end{equation}
\begin{equation}
\vec{\nabla}\times\vec{B}=\mu_0\left(\vec{J}+\vec{\nabla}\times\vec{M}+\frac{\partial\vec{P}}{\partial t}\right)+\frac{1}{c^2}\frac{\partial
\vec{E}}{\partial t}\,,
\end{equation}
and
\begin{equation}
\vec{\nabla}\cdot\vec{B}=0\,,
\end{equation}
where
\begin{equation}\label{1}
\vec{\nabla}\cdot\vec{P}=\lambda\,\vec{A}\cdot\vec{B}\,,
\end{equation}
\begin{equation}\label{1}
\vec{\nabla}\times\vec{M}+\frac{\partial\vec{P}}{\partial t}=-\lambda\left(\,\varphi\, \vec{B}-\vec{A}\times\vec{E}\right)\,,
\end{equation}
\begin{equation}
0=\varphi \,\vec{E}+c^2\vec{A}\times\vec{B}\,,
\end{equation}
and
\begin{equation}
0=\vec{A}\cdot\vec{E}\,,\qquad\qquad \lambda=\frac{qc\varepsilon_0}{\hbar}\,.
\end{equation}
Taking the dot product of $\vec{A}$ with (105) and using (104) yield the conservation equation
\begin{equation}\label{1}
\vec{\nabla}\cdot\left(\varphi\,\vec{P}-\vec{A}\times\vec{M}\right)+\frac{\partial}{\partial t}\,(\vec{A}\cdot\vec{P})=- \vec{E}\cdot\vec{P}-\vec{B}\cdot\vec{M}\,.
\end{equation}
If we now let $\vec{M}=0$, then (108) reduces to
$$
\vec{\nabla}\cdot\left(\varphi\,\vec{P}\right)+\frac{\partial}{\partial t}\,(\vec{A}\cdot\vec{P})=- \vec{E}\cdot\vec{P}\,.
$$
The divergence of (105) using (104), (101) and (102), yield
\begin{equation}\label{1}
\varepsilon_0E^2-\frac{B^2}{\mu_0}=\varphi\,\rho-\vec{A}\cdot\vec{J}\,.
\end{equation}
This equation can be compared with (16). It can be written in a covariant form as
\begin{equation}\label{1}
\frac{1}{2\mu_0}\,F^{\mu\nu}F_{\mu\nu}=A^\sigma J_\sigma\,.
\end{equation}
It is interesting to see that (27) - (30) reduce to ordinary Maxwell's equations if
\begin{equation}\label{1}
\vec{B}=\hat{A}\times\frac{\vec{E}}{c}\,, \qquad\qquad \vec{E}=c\hat{A}\times\vec{B}\,.
\end{equation}
Hence,  the three vectors, $\vec{A}$, $\vec{E}$ and $\vec{B}$ are mutually orthogonal, and that $\varphi=cA$.

\section{\textcolor[rgb]{0.00,0.07,1.00}{Concluding remarks}}

We have extended the biquaternionic Dirac's equation to include interactions with photons. The interactions of the vector and scalar potentials of the photon with the inertial and magnetic  fields made the derived Maxwell-like equations to deviate from the free ones. The magnetic mass and current densities arising from the photon interactions with the matter are found to preserve the magnetic mass. The concept of magnetic mass is introduced but no further definition is given. The axions electrodynamics developed by Frank Wilczek is found to be a special case of the present electrodynamics.   In this case the temporal and spatial variations of the axions field are related to the scalar and vector potentials of the photon, respectively.

Axion fields are found to give rise to a residual magnetic field (a Schwinger critical field type) when interacting with the charged particles. Massive axions are found to be analogous to massive photons. The conservation of electric charge is related to the direction in which the electric and magnetic fields make inside the medium. However, the total charge of the system is conserved. The electrodynamics equations  in free space are invariant under duality transformation. We have shown that the electromagnetic field vector, $\vec{F}$,  describing massive photons satisfies Dirac's equation with spin-1 matrices so that $\vec{\nabla}\cdot\vec{F}=0$.

Moreover, the electrodynamics equations we obtained are shown to express the electromagnetic fields in addition to the electric and magnetic fields developed by the existing charges in the region. The  chiral magnetic effect is shown to be associated with the additional current appearing  in Ampere's equation that is proportional to the magnetic field.

It is shown that axions  find applications in cosmology particulary it can provide significant contribution to dark matter bewildering astronomers. Besides, phenomena like topological insulator and chiral magnetic effect appearing in the realm of condensed matter can be rigourous investigated in the framework of the present electrodynamics. We undertake to pursue these issues.

\section{Appendix A: Alternative expression}

Equations (27) - (30) can be written as
\begin{equation}
\vec{\tilde{\nabla}}\times\vec{E}+\frac{\partial\vec{B}}{\partial\tau}=0\,, \qquad \vec{\tilde{\nabla}}\times\vec{B}-\frac{1}{c^2}\,\frac{\partial\vec{E}}{\partial\tau}=\mu_0\vec{J}
\end{equation}
where the operators
\begin{equation}
\vec{\tilde{\nabla}}\times=\vec{\nabla}\times+\frac{q\varphi_g}{\hbar\,c}\,,\qquad\qquad\frac{\partial}{\partial\tau}=\frac{\partial}{\partial t}+\vec{\omega}\times\,,\qquad\qquad \vec{\omega}=\frac{qc}{\hbar}\,\vec{A}_g\,,
\end{equation}
replaces the usual curl operator.
Hence, the above definition suggests that  Ampere's and Faraday's equations are expressed in a non-inertial frame (rotation frame) having angular velocity $\vec{\omega}$.
\section{Appendix B: Modified Maxwell's equations}
If we now let $\Lambda=\beta\psi_0+c\vec{\gamma}\cdot\vec{\psi}$ and abandon the condition in (10), then (5) - (8) become
\begin{equation}\label{1}
\vec{\nabla}\cdot\vec{E}_D=\frac{\rho_D}{\varepsilon_0}-\frac{qc}{\hbar}\vec{A}_g\cdot\vec{B}_D-\frac{q}{\hbar}\varphi_g\Lambda\,,
\end{equation}
\begin{equation}
\vec{\nabla}\cdot\vec{B}_D=\frac{q}{\hbar c}\vec{A}_g\cdot\vec{E}_D+\frac{1}{c}\frac{\partial \Lambda}{\partial t}\,,
\end{equation}
and
\begin{equation}\label{1}
\vec{\nabla}\times\vec{E}_D=-\frac{\partial \vec{B}_D}{\partial
t}-\frac{q}{\hbar c}\,\varphi_g \,\vec{E}_D-\frac{qc}{\hbar}\vec{A}_g\times\vec{B}_D+c\vec{\nabla}\Lambda\,,
\end{equation}
\begin{equation}
\vec{\nabla}\times\vec{B}_D=\mu_0\vec{J}_D+\frac{1}{c^2}\frac{\partial
\vec{E}_D}{\partial t}-\frac{q}{\hbar c}\,\varphi_g\, \vec{B}_D+\frac{q}{\hbar c}\vec{A}_g\times\vec{E}_D-\frac{q}{\hbar}\vec{A}_g\Lambda\,,
\end{equation}
Equations (114) - (117) are also valid for the electromagnetic field by dropping the subscript D from all terms. Let us consider the particular, case $\vec{A}_g=0$ and $\varphi_g=qV=mc^2$, and set $\nu=mc^2/\hbar$\,.  Equations (114) - (117) reduce to
\begin{equation}\label{1}
\vec{\nabla}\cdot\vec{E}=\frac{\rho}{\varepsilon_0}-\nu\,\Lambda\,,
\end{equation}
\begin{equation}
\vec{\nabla}\cdot\vec{B}=\frac{1}{c}\frac{\partial \Lambda}{\partial t}\,,
\end{equation}
and
\begin{equation}\label{1}
\vec{\nabla}\times\vec{E}=-\frac{\partial \vec{B}}{\partial
t}-\frac{\nu}{c} \,\vec{E}+c\vec{\nabla}\Lambda\,,
\end{equation}
\begin{equation}
\vec{\nabla}\times\vec{B}=\mu_0\vec{J}+\frac{1}{c^2}\frac{\partial
\vec{E}}{\partial t}-\frac{\nu}{c}\, \vec{B}\,.
\end{equation}
The energy conservation of associated with Eqs.(118) - (121) is
\begin{equation}
\vec{\nabla}\cdot\vec{S}_\Lambda+\frac{\partial u_\Lambda}{\partial t}=-\vec{E}\cdot\vec{J}\,,\qquad\qquad u_\Lambda=\frac{\varepsilon_0}{2}\,E^2+\frac{B^2}{2\mu_0}+\frac{\Lambda^2}{2\mu_0}\,,\qquad \vec{S}_\Lambda=\frac{\vec{E}\times\vec{B}-c\Lambda\vec{B}}{\mu_0}\,.
\end{equation}
The above equation shows that some of energy flows along the magnetic field direction. In addition, $\Lambda$ is a real physical field having energy and momentum that doesn't interact with the electric current.
It is interesting to observe that a magnetic wave with a vanishing electric field is possible, as evident from Eq.(122). This wave flows along the magnetic field direction, which doesn't experience any dissipation ($\vec{E}\cdot\vec{J}=0$). We should therefore search for such a wave. Furthermore, Eqs.(118) and (119) show that magnetic energy density is non-zero when a time varying scalar filed $\Lambda$ is present. A magnetic current density arises when the scalar filed has a non-uniform distribution, \emph{viz}., $\vec{J}_m=-c\vec{\nabla}\Lambda$. The electric charge conservation can be obtained from Eqs.(118), (119) and (121) as
\begin{equation}
\vec{\nabla}\cdot\vec{J}+\frac{\partial \rho_{eff.}}{\partial t}=0\,,\qquad\qquad \rho_{eff.}=\rho-2\nu\varepsilon_0\Lambda\,.
\end{equation}
The above equation reduces to the standard form upon setting $\Lambda$ or $m$ ($\nu$) to zero.

\end{document}